\documentclass[runningheads]{llncs}

\usepackage[T1]{fontenc}
\usepackage{textcomp}
\usepackage{lmodern}
\usepackage{graphicx,booktabs,tabularx,xcolor,hyperref}
\usepackage{amsmath,subcaption,float,amsfonts,amssymb,bm,mathtools,booktabs,threeparttable,nccmath}
\usepackage[title]{appendix}
\hypersetup{colorlinks=true,linkcolor=[rgb]{0.1,0.3,0.9},urlcolor=[rgb]{0.2,0.2,0.2},citecolor=[rgb]{0.1,0.3,0.9}}
\usepackage{float}

\begin{document}

\author{Simon Winther Albertsen \and Hjalte Svaneborg Bjørnstrup \and \\ Mostafa Mehdipour Ghazi \thanks{Corresponding author: ghazi@di.ku.dk}}
\authorrunning{S. Winther et al.}
\institute{Pioneer Centre for Artificial Intelligence, Department of Computer Science, University of Copenhagen, Copenhagen, Denmark}

\title{RARE-UNet: Resolution-Aligned Routing Entry for Adaptive Medical Image Segmentation}
\titlerunning{RARE-UNet}

\maketitle

\begin{abstract}

Accurate segmentation is crucial for clinical applications, but existing models often assume fixed, high-resolution inputs and degrade significantly when faced with lower-resolution data in real-world scenarios. To address this limitation, we propose RARE-UNet, a resolution-aware multi-scale segmentation architecture that dynamically adapts its inference path to the spatial resolution of the input. Central to our design are multi-scale blocks integrated at multiple encoder depths, a resolution-aware routing mechanism, and consistency-driven training that aligns multi-resolution features with full-resolution representations. We evaluate RARE-UNet on two benchmark brain imaging tasks for hippocampus and tumor segmentation. Compared to standard UNet, its multi-resolution augmented variant, and nnUNet, our model achieves the highest average Dice scores of 0.84 and 0.65 across resolution, while maintaining consistent performance and significantly reduced inference time at lower resolutions. These results highlight the effectiveness and scalability of our architecture in achieving resolution-robust segmentation. The codes are available at: \url{https://github.com/simonsejse/RARE-UNet}.

.

\end{abstract}

\section{Introduction}

Accurate segmentation in magnetic resonance imaging (MRI) is vital for clinical workflows, supporting tasks from developmental studies to diagnosis and treatment planning. As multi-institutional neuroimaging efforts grow \cite{jack2024overview,marcus2007open}, there is an increasing need for robust segmentation methods that generalize across heterogeneous MRI scans with varying acquisition protocols. Yet, most deep learning models, especially UNet variants \cite{ronneberger2015u,zhou2018unet++,zhou2019unet++,huang2020unet,ibtehaz2020multiresunet,azad2024medical}, assume fixed input resolution and spatial shape, an assumption rarely valid in real-world clinical data.

In practice, resolution and quality vary due to differences in scanners, protocols, and patient-specific factors. This variability is common in real-world clinical settings and large-scale studies, where scans are acquired under non-standardized conditions and often exhibit inconsistent resolution, noise levels, or structural clarity. Multi-center datasets introduce variability due to diverse voxel spacing and anisotropic resolutions. Standard pipelines often rely on pre-processing to enforce shape consistency, but such operations can degrade image quality, e.g., resampling may lose small structures or introduce blur or artifacts. Over-padding can also skew predictions toward background regions.

We propose RARE-UNet, a resolution-adaptive extension of the UNet tailored for efficient segmentation of variable-resolution brain MRI. The core idea is a set of multi-scale blocks (MSBs), which allow inputs to enter the encoder at depths matched to their spatial resolution. While prior multi-input multi-scale networks \cite{cho2021rethinking,lin2023image} have been successfully employed in image restoration tasks, their focus is on fusing visual features for improved quality. Moreover, multi-scale segmentation networks, designed to enhance receptive field diversity, do not process multi-resolution inputs \cite{chen2017deeplab}. In contrast, our approach is designed for semantic segmentation and introduces a resolution-aligned entry mechanism within a single UNet architecture, reusing features across scales. Rather than processing all inputs through full or parallel branches, our model adapts dynamically to input resolution, enabling only the semantically relevant network pathway, and enforces deep supervision and cross-scale consistency for robust anatomical delineation. Low-resolution inputs are injected directly into deeper layers to leverage semantic representations, optimizing computational efficiency.

This approach eliminates the need for heavy pre-processing and aligns feature abstraction levels with input fidelity. During inference, a lightweight resolution-based routing mechanism selects the appropriate MSB, enabling efficient, adaptive processing. Our contributions are as follows: (1) We introduce a novel UNet variant with resolution-aware MSBs for resolution-adaptive segmentation using a single shared architecture. (2) We propose a deep supervision strategy where each resolution path contributes to the loss, combined with a scale consistency loss to align features across scales. (3) We present an efficient inference-time routing mechanism that activates only relevant encoder layers based on input resolution, reducing preprocessing and compute costs. (4) We conduct extensive experiments on 3D brain MRI datasets showing improved segmentation accuracy and robustness across diverse resolutions, while offering significant computational savings and outperforming baseline UNets and nnUNet \cite{isensee2021nnu}.

\section{Related Work}

UNet \cite{ronneberger2015u} is a widely used architecture for medical image segmentation, featuring an encoder-decoder structure with skip connections. The encoder progressively downsamples the input image using convolutional blocks, typically composed of convolutional layers, batch normalization, and ReLU activation, until reaching a bottleneck representation. The decoder then upsamples these features back to the input resolution, fusing encoder features at each level via skip connections. These connections preserve spatial detail and enable better gradient flow. In its original form, UNet applies supervision only at the final output layer, comparing the prediction with the ground truth.

Several extensions have been proposed to enhance UNet’s capacity for multi-scale representation and deeper supervision. UNet+ \cite{zhou2018unet++} introduces intermediate decoder branches for each encoder level, enabling supervision at multiple output paths. Each path independently contributes to the final prediction, allowing the model to learn from both fine and coarse features during training. UNet++ \cite{zhou2019unet++} refines the skip connection mechanism by introducing nested and dense skip pathways that link encoder and decoder blocks across multiple levels. This design encourages feature reusability and semantic alignment, improving the model’s ability to segment objects with varying sizes and boundaries.

UNet 3+ \cite{huang2020unet} further generalizes this concept by integrating full-scale skip connections across all encoder and decoder stages. At each decoder level, features from every resolution are aggregated, enabling the model to learn a comprehensive representation that captures both local and global context. Deep supervision is applied throughout the network, strengthening gradient propagation and multiscale learning. This technique introduces auxiliary outputs at multiple decoder stages, each guided by its own loss function, encouraging the network to learn meaningful representations at different semantic depths.

In addition to these architectural advances, other techniques such as pyramid pooling and dilated atrous convolutions have been employed to enhance receptive field diversity \cite{chen2017deeplab}, supporting better performance on tasks involving scale variation. Despite these improvements, most existing models still operate under the assumption of fixed input resolution and shape. They apply the full encoder path uniformly to all inputs, regardless of native resolution or quality. This leads to inefficiencies when processing low-resolution images, where early encoder layers may contribute little to the final segmentation. Moreover, such architectures lack mechanisms to adapt dynamically to input scale, limiting their generalizability in real-world settings with heterogeneous imaging data.

\section{Methods}

\subsection{RARE-UNet Architecture}

\begin{figure}[t]
\centering
\includegraphics[width=0.95\linewidth]{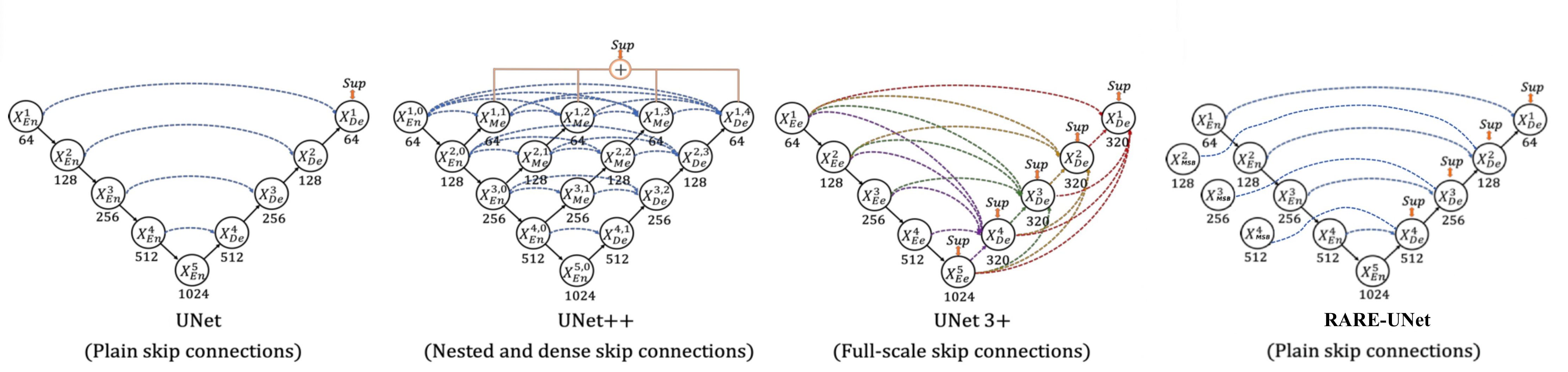}
\caption{Comparison of UNet, UNet++, UNet 3+, and the RARE-UNet architecture, highlighting key structural differences adapted from \cite{huang2020unet}. While prior models focus on dense skip connections and feature-level fusion, RARE-UNet introduces a simpler yet more efficient design based on resolution-aware input routing. This enables scale-adaptive processing without duplicating encoder paths.}
\label{fig:unet_variants}
\end{figure}

\begin{figure}[!t]
\centering
\includegraphics[width=0.95\linewidth]{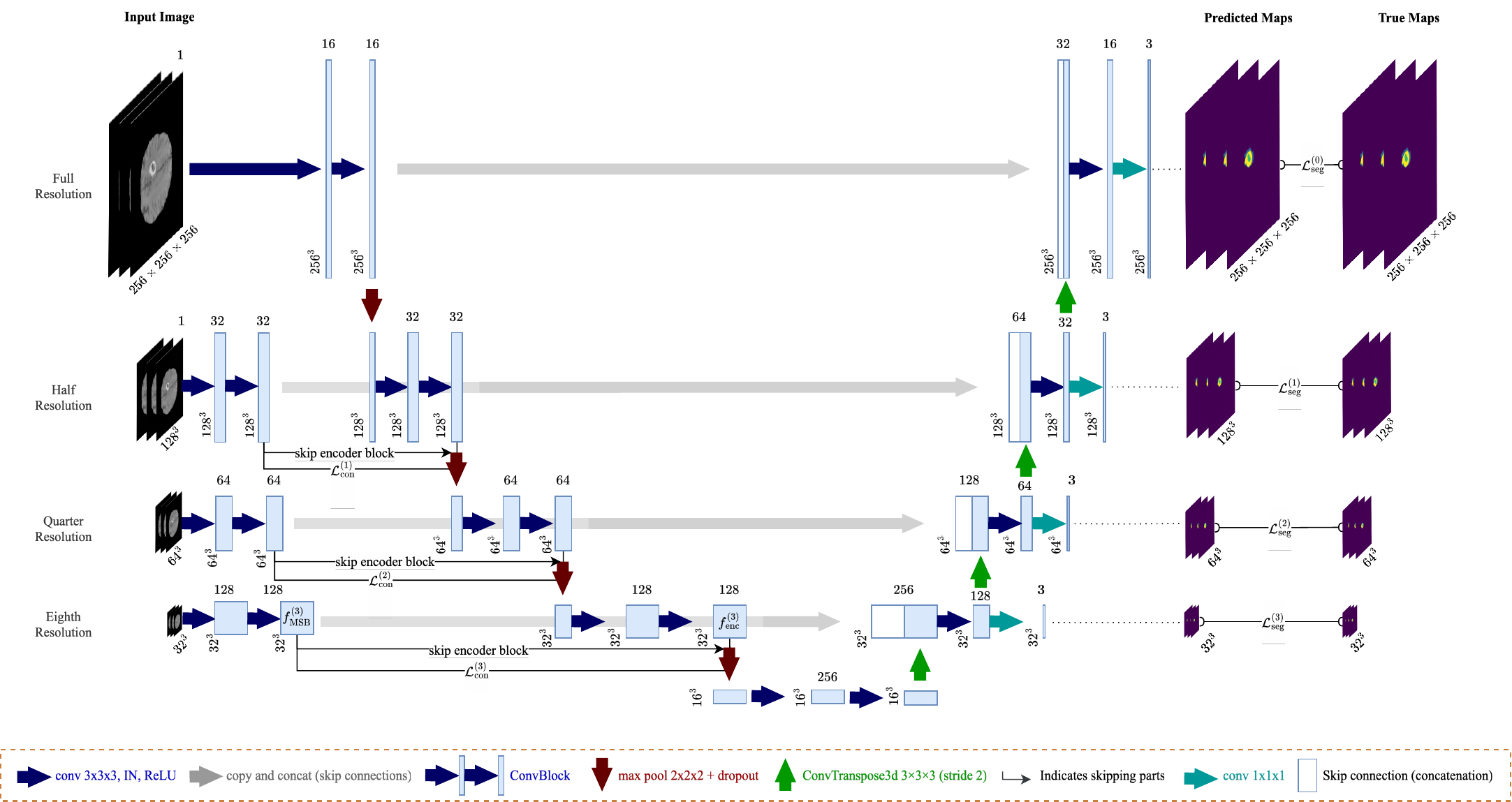}
\caption{Overview of the RARE-UNet architecture. The model extends a standard 3D UNet with multi-scale blocks, serving as resolution-aware entry points at various encoder depths. Full-resolution inputs are processed from the first encoder layer, while low-resolution inputs (e.g., $1/2$, $1/4$, $1/8$ scale) are routed to deeper blocks. Each path shares the same bottleneck and proceeds through a resolution-aligned decoder with dedicated segmentation heads. The example shown uses four resolution levels, but the architecture can be scaled deeper or shallower depending on the input size and available computational resources.}
\label{fig:Architechture}
\end{figure}

Fig. \ref{fig:unet_variants} contrasts our resolution-aligned routing design with traditional multi-scale architectures like UNet++ and UNet 3+, which rely on dense skip connections and feature-level fusion across fixed-resolution inputs. Unlike prior multi-scale designs that primarily fuse features at the decoder stage, RARE-UNet introduces scale-awareness directly at the input level. As shown in Fig. \ref{fig:Architechture}, our model is built upon a standard 3D UNet backbone with an encoder-decoder structure, skip connections, and a shared bottleneck. The core innovation lies in the introduction of multi-scale gateway blocks, serving as resolution-aware routing entry points at different encoder depths. This design enables the model to accept both full- and downsampled-resolution inputs routed to appropriate depths without the need for global resampling or shape adjustment.

For example, a full-resolution input enters the network at the first encoder layer (depth 0), while coarser inputs (e.g., $1/2$, $1/4$, or $1/8$ scale) bypass the early encoders and are injected directly via MSBs at deeper layers. This routing strategy not only preserves image fidelity but also avoids redundant computations for low-resolution scans. Each resolution path has its own dedicated segmentation head, allowing for independent predictions and supervision. RARE-UNet thus provides a unified and efficient framework for processing variable-resolution inputs through a shared encoder-decoder network, adapting computation dynamically to input scale while retaining the structural advantages of UNet.

\subsection{Multi-Scale Gateway Blocks}

\begin{figure}[t]
\centering
\includegraphics[width=0.825\linewidth]{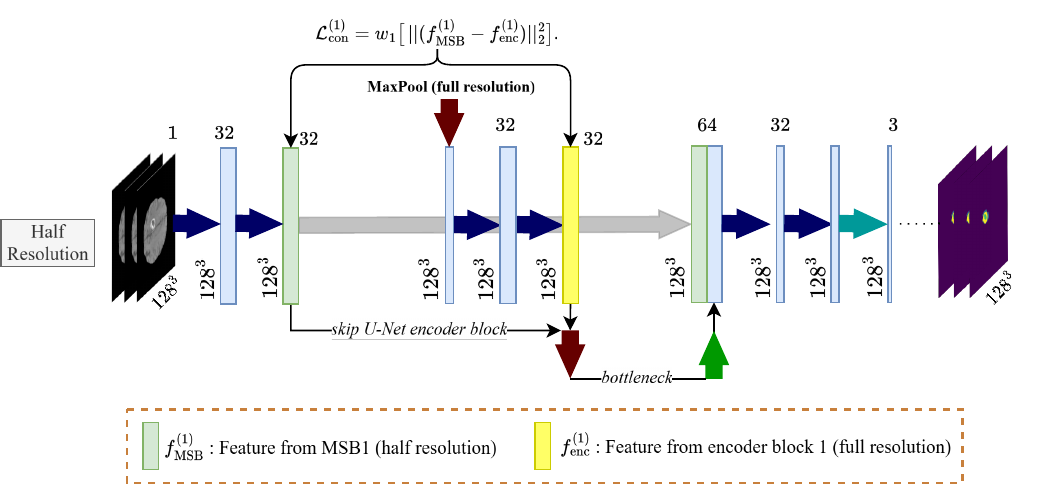}
\caption{Illustration of a Multi-Scale Gateway Block at depth 1. A downsampled input ($1/2$ resolution) enters the network via MSB1, which generates features aligned with the encoder output at that depth. These features are used both as skip connections and as input to deeper layers. A dedicated segmentation head produces the output for this resolution.}
\label{fig:Architechture_part}
\end{figure}

The gateway blocks enable resolution-adaptive processing by routing inputs at different scales into the UNet at corresponding encoder depths. Fig. \ref{fig:Architechture_part} shows a multi-scale gateway block at depth 1 (MSB1). A $1/2$-resolution input enters at depth 1 via MSB1, which transforms it into a feature map $f_{\mathrm{MSB}}^{(1)}$ that aligns in shape and semantics with the standard encoder output $f_{\mathrm{enc}}^{(1)}$. To ensure this alignment, we apply a mean squared error (MSE) consistency loss between these representations during training. The MSB output is then used both as the skip connection and as the input to deeper layers (from depth 2 onward). A resolution-specific segmentation head (a $1 \times 1 \times 1$ convolution) generates the prediction for this path, in contrast to the shared final head used by the full-resolution stream.

\paragraph{\textbf{Training.}}

During training, inputs are downsampled by factors of $2^d$ to create lower-resolution versions. Each version follows a path through the network appropriate to its resolution. The full-resolution input is processed through the entire encoder-decoder pipeline, storing intermediate encoder outputs as targets for training MSBs. Each downsampled input enters via its corresponding MSB (e.g., MSB2 for $1/4$ scale), which generates features that approximate what the encoder would produce at that depth. These features are passed through the remaining encoder layers, the bottleneck, and decoder blocks up to their resolution level. The segmentation loss is computed separately for each output using resolution-specific heads. This setup allows the model to learn consistent representations across scales, enabling it to generalize to varying input resolutions without requiring resampling, separate models, or fusion schemes.

\paragraph{\textbf{Inference.}}

At inference time, the input image size is rounded to the nearest matching resolution level. Full-resolution inputs follow the standard encoder path; low-resolution inputs are routed through the corresponding MSBs, skipping shallower encoder and decoder layers. The segmentation is produced using the head specific to that resolution. This resolution-aware routing reduces unnecessary computation and allows the model to adapt to a range of input sizes without modification to the network or input shape.

\subsection{Loss Function}

\paragraph{\textbf{Segmentation Loss.}}

Each resolution path produces a segmentation output, compared against a downsampled version of the ground truth using nearest-neighbor interpolation to preserve label structure. We compute a weighted combination of cross-entropy and soft Dice loss at each depth $d$, given by $\mathcal{L}_{\mathrm{seg}}^{(d)} = 
\alpha\,\mathcal{L}_{\mathrm{CE}}^{(d)} + (1 - \alpha)\,\mathcal{L}_{\mathrm{Dice}}^{(d)}$, where $\alpha \in [0,1]$ is initially set to 0.5 in our experiments. The total segmentation loss is the average over active resolution paths.

\paragraph{\textbf{Consistency Loss.}}

To align feature representations across scales, we introduce a consistency loss between the MSB output $f_{\mathrm{MSB}}^{(d)}$ and the corresponding encoder feature $f_{\mathrm{enc}}^{(d)}$ using MSE: $\mathcal{L}_{\mathrm{con}}^{(d)} = \lVert f_{\mathrm{MSB}}^{(d)} - f_{\mathrm{enc}}^{(d)} \rVert_2^2$. These are averaged across all MSB levels to obtain $\mathcal{L}_{\mathrm{con}}$. The final training loss is a weighted sum of segmentation and consistency terms: $\mathcal{L}_{\mathrm{total}} = \mathcal{L}_{\mathrm{seg}} + \lambda_{\mathrm{con}} \cdot \mathcal{L}_{\mathrm{con}}$, where the regularization factor $\lambda_{\mathrm{con}}$ is initially set to 1 in our experiments.

\section{Experiments and Results}

\subsection{Data}

We evaluate our model on two publicly available brain imaging datasets: BraTS \cite{bakas2017advancing,bakas2018identifying} (484 labeled 4D scans) and VUMC \cite{simpson2019large} (260 labeled 3D scans). These datasets were preprocessed and resampled to 1 mm$^3$ isotropic resolution as part of the Medical Segmentation Decathlon \cite{antonelli2022medical}, corresponding to the first and fourth tasks (brain tumor and hippocampus segmentation). For hippocampus segmentation, volumes were cropped or padded to a fixed size of 32$\times$64$\times$32 voxels. For tumor segmentation, we used 256$\times$256$\times$128 volumes. Each dataset was randomly split into 80\% training/validation and 20\% testing. Intensity normalization involved clipping voxel values to the [0.5, 99.5] percentile range and stretching the contrast in the [0, 1] range. A detailed description of both datasets and pre-processing steps is provided in the Appendix.

\subsection{Experimental Setup}

We evaluate RARE-UNet against three baselines: (1) a standard 3D UNet, (2) a UNet trained with multi-resolution input augmentation (UNet+Aug), and (3) nnUNet, a state-of-the-art self-configuring framework that automatically adapts pre-processing, augmentation, patch size, batch size, and other training settings to each dataset. To ensure a fair and controlled comparison focused solely on architectural and algorithmic differences, we disabled optional features such as deep supervision and aggressive data augmentation in all UNet variants. All models share similar architectural choices, unless otherwise specified.

To isolate the effect of our multi-scale architecture from that of resolution diversity during training, we implemented the UNet+Aug variant, which randomly receives either full-resolution or downsampled inputs with a 50\% chance. The downsampling factor is randomly selected from ${1/2, 1/4, 1/8}$, corresponding to different encoder depths. These inputs are then zero-padded or upsampled to the original resolution, ensuring consistency with the ground-truths and preventing the model from learning to simulate resolution variation through pre-processing.

Network parameters were optimized using the AdamW optimizer. To identify the best configuration, we performed an extensive sweep of over 100 runs, varying the learning rate and consistency loss weight. The optimal setup was selected based on validation performance averaged across all resolution scales. It should be noted that we applied equal weighting to the loss contributions from all resolution scales to ensure balanced performance across inputs of varying sizes, rather than optimizing for full-resolution at the expense of others.

We assess segmentation accuracy using the Dice Similarity Coefficient (DSC). We used Weights \& Biases \cite{wandb2020} to track training and validation metrics, including loss functions and segmentation performance across scales. To simulate multi-resolution inputs, we downsample each scan using linear interpolation at scales of $1/2$, $1/4$, and $1/8$ of the original resolution. For UNet-based models, low-resolution inputs are either padded (Pad) or upsampled (Up) to match the original resolution, consistent with their respective training protocols. For each dataset and model, we report DSC at the full resolution and each downsampled scale, along with the average DSC across all resolutions.

\subsection{Results}

\paragraph{\textbf{Hippocampus Segmentation.}}

Table \ref{tab:mean-dsc-hp} reports the Dice scores on the hippocampus segmentation test set. Our RARE-UNet consistently outperforms all baselines across downsampled resolutions, while remaining competitive with state-of-the-art models at full resolution. As expected, UNet variants with upsampling generally outperform those with zero-padding when handling low-resolution inputs. Besides, nnUNet achieves the highest DSC at full resolution, consistent with its extensive auto-configuration and preprocessing pipeline.

\begin{table}[t]
\centering
\footnotesize
\caption{Comparison of test DSCs (mean $\pm$ SD) for hippocampus segmentation at multiple resolutions. The best results per resolution are highlighted. The proposed model achieves high accuracy and low variance across scales.}
\vspace{0.1cm}
\label{tab:mean-dsc-hp}
\resizebox{\textwidth}{!}{
\begin{tabular}{lccccc}
\toprule
\textbf{Method}
& \multicolumn{1}{c}{\textbf{Scale 1}}
& \multicolumn{1}{c}{\textbf{Scale 1/2}}
& \multicolumn{1}{c}{\textbf{Scale 1/4}}
& \multicolumn{1}{c}{\textbf{Scale 1/8}}
& \multicolumn{1}{c}{\textbf{Overall}} \\
\midrule
UNet--Pad & \underline{0.874}$\pm$0.037 & 0.034$\pm$0.033 & 0.000$\pm$0.000 & 0.000$\pm$0.000 & 0.227$\pm$0.374 \\
UNet--Up & \underline{0.874}$\pm$0.037 & 0.848$\pm$0.058 & 0.776$\pm$0.083 & 0.334$\pm$0.150 & 0.708$\pm$0.219 \\
\midrule
UNet+Aug--Pad & 0.871$\pm$0.034 & 0.213$\pm$0.034 & 0.066$\pm$0.026 & 0.000$\pm$0.000 & 0.287$\pm$0.345 \\
UNet+Aug--Up & 0.871$\pm$0.034 & \underline{0.860}$\pm$0.039 & \underline{0.822}$\pm$0.069 & \textbf{0.789}$\pm$0.131 & \underline{0.835}$\pm$0.032 \\
\midrule
nnUNet--Pad & \textbf{0.880}$\pm$0.032 & 0.819$\pm$0.045 & 0.529$\pm$0.097 & 0.030$\pm$0.062 & 0.564$\pm$0.336 \\
nnUNet--Up & \textbf{0.880}$\pm$0.032 & 0.820$\pm$0.067 & 0.597$\pm$0.106 & 0.055$\pm$0.085 & 0.588$\pm$0.325 \\
\midrule
RARE-UNet & 0.869$\pm$0.032 & \textbf{0.862}$\pm$0.037 & \textbf{0.835}$\pm$0.063 & \underline{0.785}$\pm$0.134 & \textbf{0.838}$\pm$0.033 \\
\bottomrule
\end{tabular}
}
\end{table}

\begin{figure}[!t]
\centering
\includegraphics[width=0.8\textwidth]{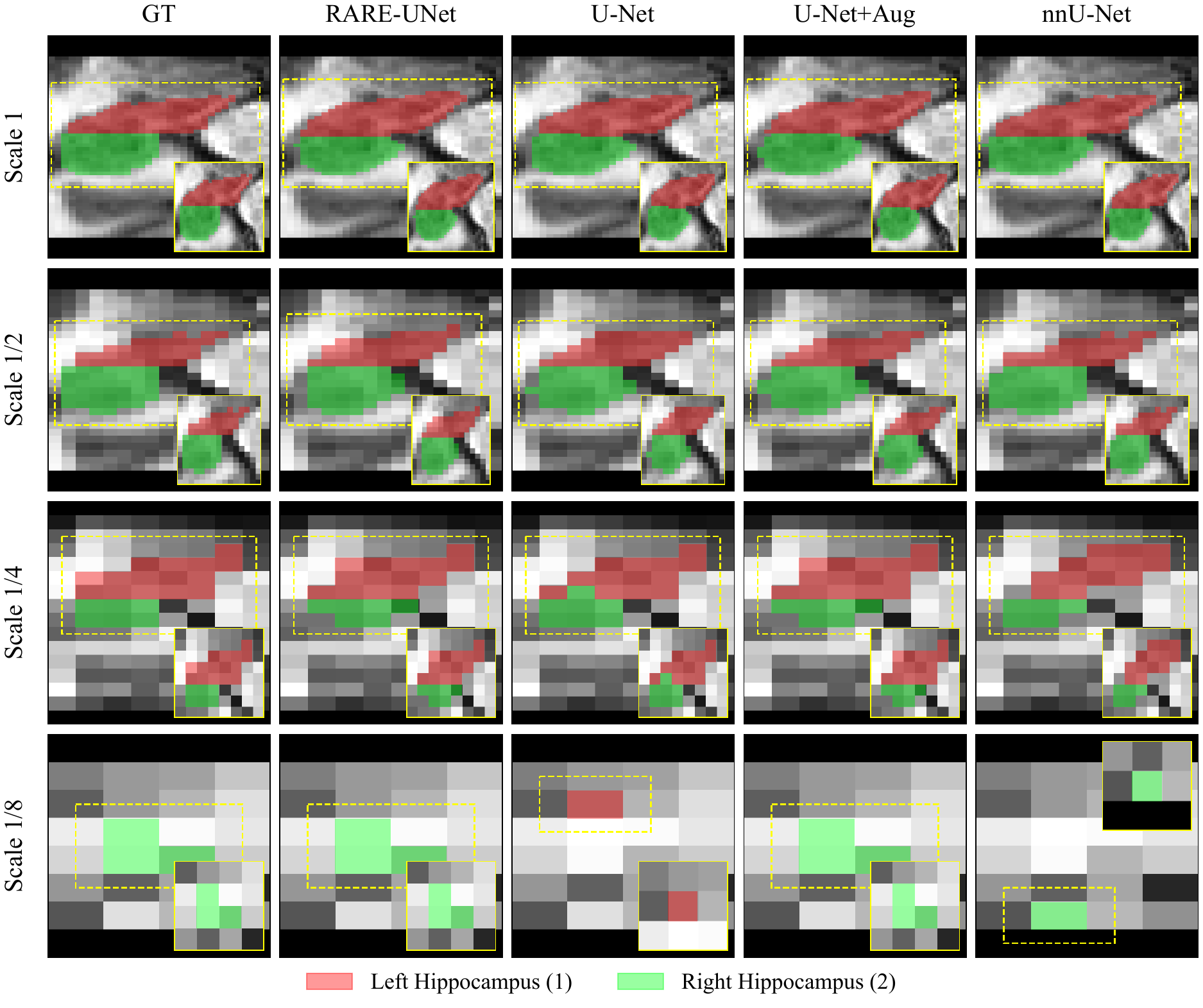}
\caption{Qualitative comparison of hippocampus segmentation results on a sample test scan. Axial slices are shown for each model to highlight differences in anatomical delineation per scale.}
\label{fig:dice_visualization_hp}
\end{figure}

More specifically, when considering average performance across all resolutions, our model achieves an average DSC of 0.84, which is approximately 0.25 higher than nnUNet, demonstrating better robustness to varying input resolutions. Additionally, RARE-UNet exhibits lower standard deviation across scales, indicating greater consistency and generalization. Notably, the performance gap becomes more pronounced at lower resolutions, where baseline models experience a significant drop in accuracy, while RARE-UNet maintains stable performance.

Interestingly, the U-Net trained with resolution augmentation and upsampling achieves performance comparable to RARE-UNet at certain scales. However, it is important to note that this approach does not yield architectural adaptivity or efficiency. While it exposes the model to resolution variability during training, it still relies on a fixed architecture and incurs the same computational cost regardless of input resolution. In contrast, RARE-UNet offers dynamic resolution-aware processing with reduced complexity for lower-resolution.

These results highlight the scalability and resolution-adaptive capabilities of our architecture for volumetric medical image segmentation. Qualitative results are shown in Fig. \ref{fig:dice_visualization_hp}, where segmentation outputs are visualized for a sample test scan. The differences between methods are subtle at full resolution, though the proposed model better preserves structural detail across resolutions. Detailed per-class segmentation accuracies are provided in the Appendix.

\paragraph{\textbf{Brain Tumor Segmentation.}}

Table \ref{tab:mean-dsc-bt} reports the DSCs on the brain tumor test set across multiple resolutions. As expected, nnUNet achieves the highest accuracy at full resolution (0.71), followed closely by UNet and our proposed RARE-UNet (0.69). However, at lower resolutions, RARE-UNet consistently outperforms all baselines. Notably, our model achieves the highest average DSC across all scales (0.65), surpassing nnUNet (0.61) and other alternatives. Additionally, RARE-UNet exhibits the smallest variation in performance across resolutions, demonstrating greater stability and robustness to input scale.

\begin{table}[b]
\centering
\footnotesize
\caption{Comparison of test DSCs (mean $\pm$ SD) for brain tumor segmentation at multiple resolutions. The best results per resolution are highlighted. The proposed model achieves high accuracy and low variance across scales.}
\vspace{0.1cm}
\label{tab:mean-dsc-bt}
\resizebox{\textwidth}{!}{
\begin{tabular}{lccccc}
\toprule
\textbf{Method}
& \multicolumn{1}{c}{\textbf{Scale 1}}
& \multicolumn{1}{c}{\textbf{Scale 1/2}}
& \multicolumn{1}{c}{\textbf{Scale 1/4}}
& \multicolumn{1}{c}{\textbf{Scale 1/8}}
& \multicolumn{1}{c}{\textbf{Overall}} \\
\midrule
UNet--Pad & \underline{0.699}$\pm$0.154 & 0.208$\pm$0.124 & 0.176$\pm$0.114 & 0.066$\pm$0.062 & 0.287$\pm$0.244 \\
UNet--Up & \underline{0.699}$\pm$0.154 & \underline{0.677}$\pm$0.151 & 0.620$\pm$0.133 & 0.493$\pm$0.134 & 0.622$\pm$0.080 \\\midrule
UNet+Aug--Pad & 0.688$\pm$0.144 & 0.072$\pm$0.058 & 0.033$\pm$0.038 & 0.046$\pm$0.094 & 0.210$\pm$0.277 \\
UNet+Aug--Up & 0.688$\pm$0.144 & 0.673$\pm$0.139 & \underline{0.627}$\pm$0.118 & \underline{0.509}$\pm$0.119 & \underline{0.624}$\pm$0.070 \\\midrule
nnUNet--Pad & \textbf{0.712}$\pm$0.152 & 0.670$\pm$0.152 & 0.583$\pm$0.128 & 0.399$\pm$0.125 & 0.591$\pm$0.120 \\
nnUNet--Up & \textbf{0.712}$\pm$0.152 & 0.660$\pm$0.155 & 0.595$\pm$0.140 & 0.452$\pm$0.145 & 0.605$\pm$0.097 \\\midrule
RARE-UNet & 0.693$\pm$0.159 & \textbf{0.693}$\pm$0.154 & \textbf{0.652}$\pm$0.138 & \textbf{0.567}$\pm$0.162 & \textbf{0.651}$\pm$0.052 \\
\bottomrule
\end{tabular}
}
\end{table}

\begin{figure}[!t]
\centering
\includegraphics[width=0.775\textwidth]{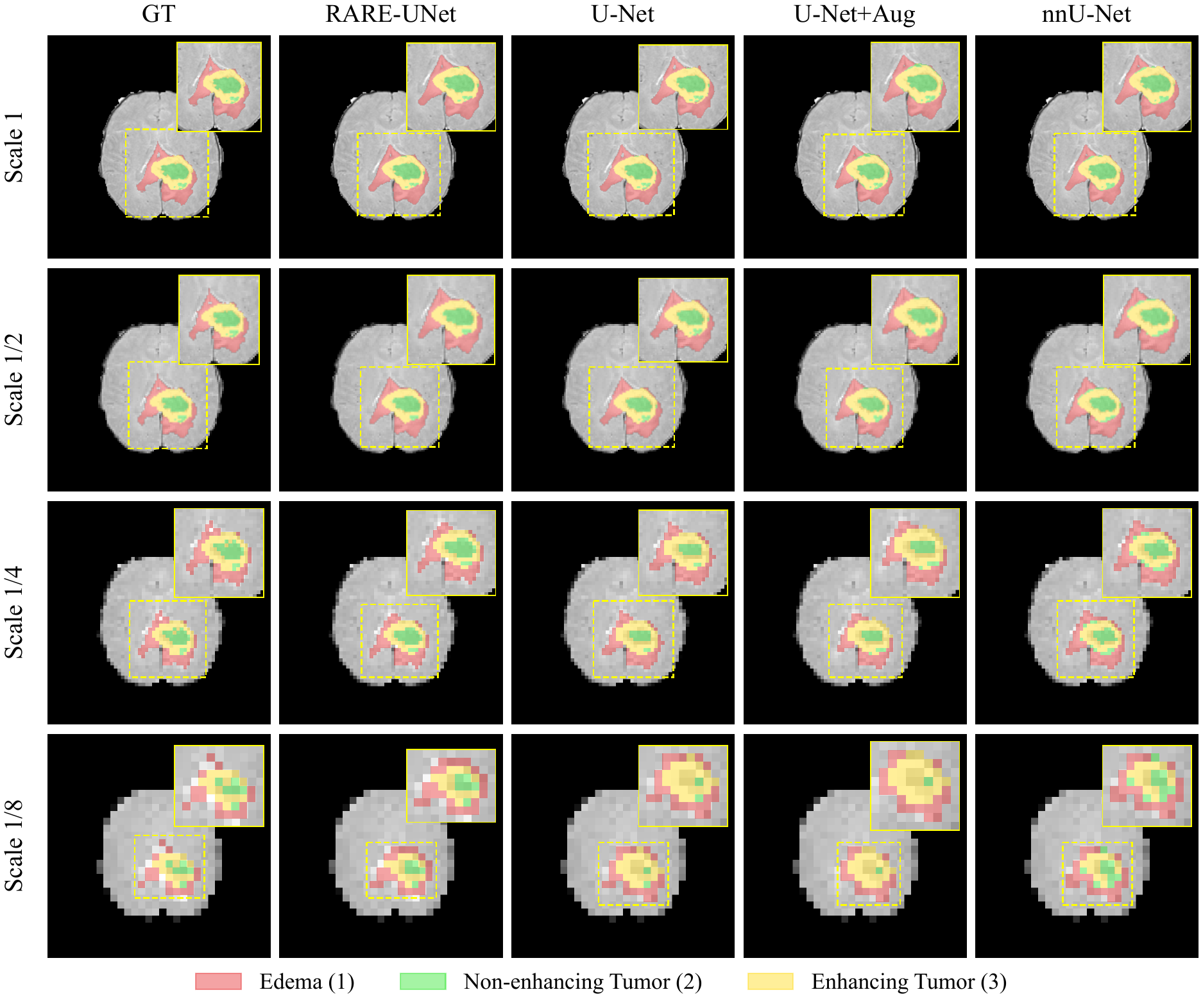}
\caption{Qualitative comparison of brain tumor segmentation results on a sample test scan. Axial slices are shown for each model to highlight differences in anatomical delineation per scale.}
\label{fig:dice_visualization_bt}
\end{figure}

The superior performance of nnUNet at full resolution may be partially attributed to its patch-based strategy, which enables more memory-efficient processing of high-resolution volumes. This design typically increases training sample diversity and enhances robustness by mitigating translational variances, factors particularly beneficial in non-symmetric and anatomically variable tasks such as tumor segmentation. However, such patch-based approaches require additional post-processing (e.g., Gaussian smoothing) to correct boundary artifacts, which may impact both efficiency and precision in multi-scale contexts. 

Fig. \ref{fig:dice_visualization_bt} provides qualitative comparisons of brain tumor segmentation results for a sample test scan. While both RARE-UNet and nnUNet produce accurate predictions, our model demonstrates superior structural consistency across tumor subregions, reinforcing its multi-scale generalization capability. More quantitative results are included in the Appendix.

\paragraph{\textbf{Model Efficiency.}}

\begin{figure}[!t]
\centering
\begin{subfigure}[b]{0.49\linewidth}
\centering
\includegraphics[width=\linewidth]{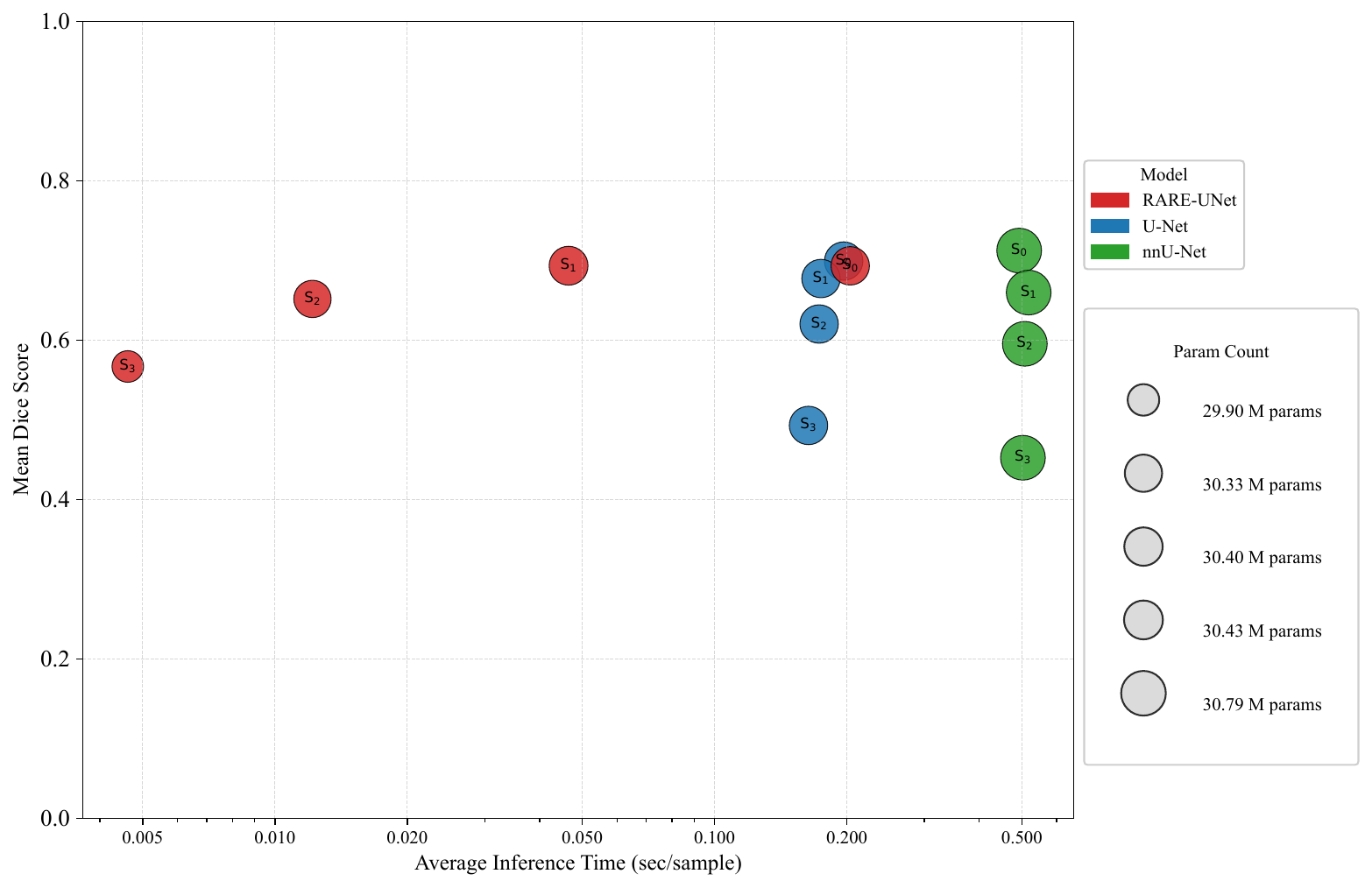}
\caption{Brain Tumor}
\end{subfigure}
\hfill
\begin{subfigure}[b]{0.49\linewidth}
\centering
\includegraphics[width=\linewidth]{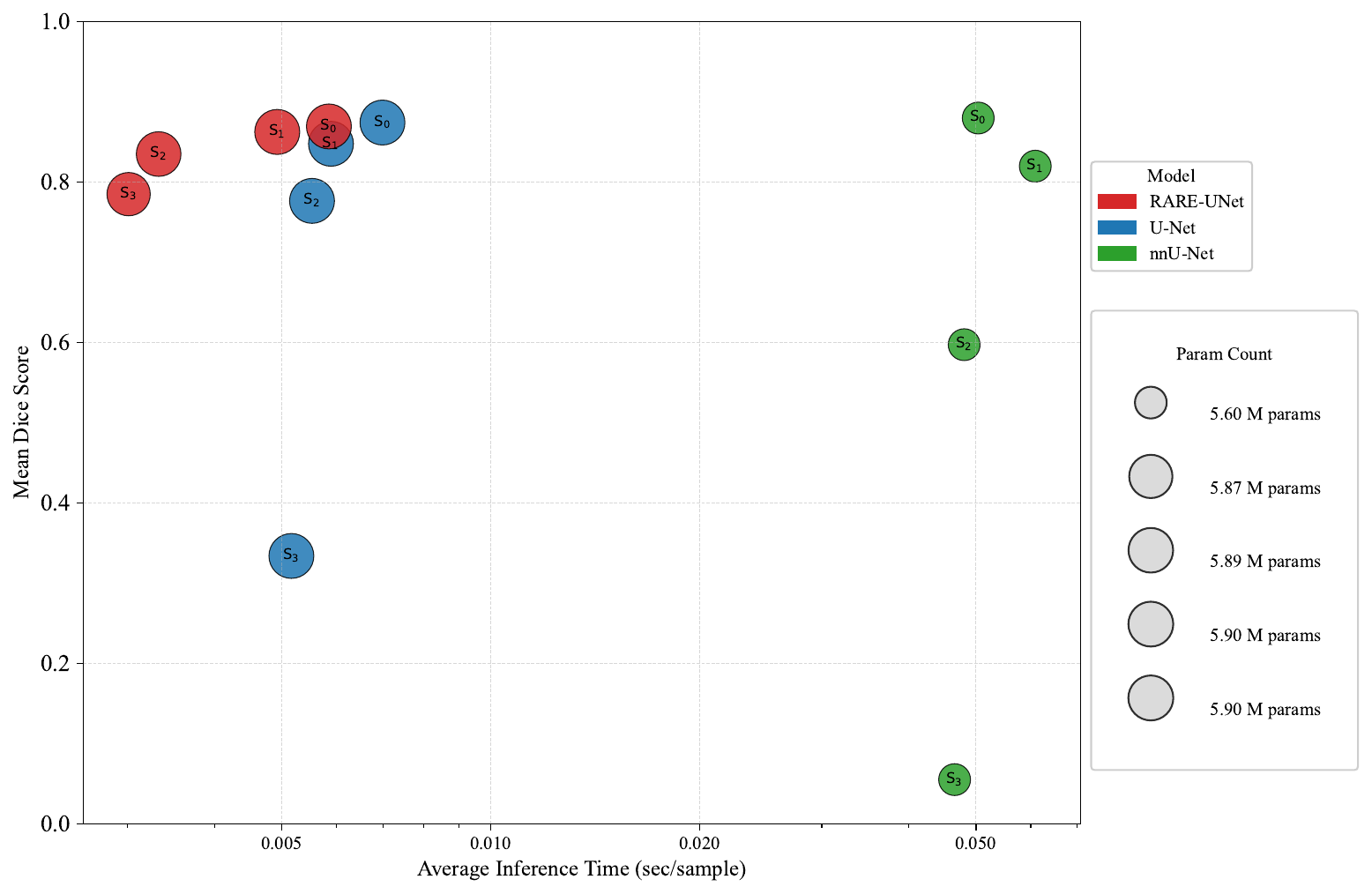}
\caption{Hippocampus}
\end{subfigure}
\caption{Trade-off between segmentation accuracy, inference time, and model complexity for different methods across multiple resolutions. For clarity, only the upsampling variants of the UNet baselines and nnUNet are shown, as they consistently outperformed their padding counterparts. S0 through S4 denote image resolutions from no scaling to scale 4 (1/8 of the original resolution).}
\label{fig:dice_vs_time}
\end{figure}

In addition to segmentation accuracy, we assess the computational efficiency of each model by measuring the average inference time over 10 runs on the test set for both tasks. Fig. \ref{fig:dice_vs_time} illustrates the trade-offs among inference time, segmentation accuracy, and model complexity (number of parameters) for the hippocampus and brain tumor segmentation tasks. Although we aligned the number of filters and parameters across models for a fair comparison, minor differences in model size persisted, likely due to architectural variations such as different convolutional blocks or implicit optimization strategies.

RARE-UNet achieves stable performance across resolutions with significantly reduced inference time, stemming from its multi-input architecture that efficiently leverages lower-resolution inputs without sacrificing quality. In brain tumor segmentation, as the spatial resolution is halved, RARE-UNet achieves a 4 times speedup without a noticeable drop in accuracy. This efficiency gain arises because halving each spatial dimension reduces the number of voxel-wise computations by a factor of $1/8$, while the number of channels doubles per encoder stage ($2 \times 1/8$), resulting in a favorable efficiency-accuracy trade-off.

\section{Conclusion}

We proposed a resolution-aware architecture for medical image segmentation, capable of maintaining high performance across a wide range of input image resolutions. Unlike standard UNets, our model incorporates multi-resolution branches supervised by both segmentation and consistency losses to promote resolution-robust feature learning. We demonstrated the effectiveness of our method on two benchmark datasets, showing that RARE-UNet achieves competitive accuracy at full resolution and outperforms state-of-the-art baselines at reduced resolutions, especially when compared to nnUNet. Our method also offers improved inference efficiency through its dynamic parameter scaling with input resolution, achieving significant speed-ups without compromising segmentation accuracy. This makes the proposed approach promising for real-world clinical workflows, where image resolution may vary due to protocols or hardware limitations. 

\section*{Acknowledgments}

This project is supported by the Pioneer Centre for AI, funded by the Danish National Research Foundation (grant number P1).

\bibliographystyle{splncs}
\bibliography{references}

\newpage
\appendix
\chapter*{Appendix}

\section{Data and Processing}

\paragraph{\textbf{Brain Tumor.}}

\begin{figure}[!b]
\centering
\begin{subfigure}[b]{0.49\linewidth}
\centering
\includegraphics[width=\linewidth]{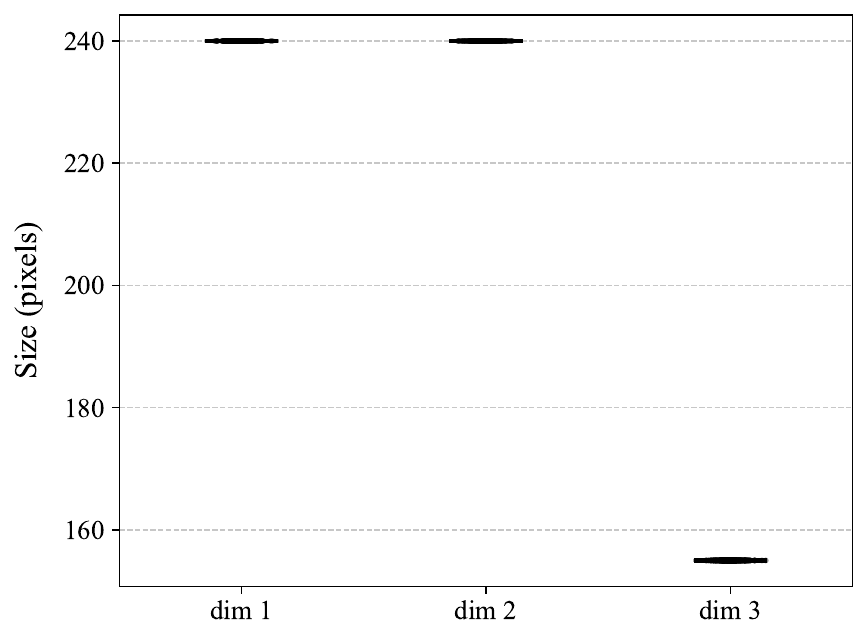}
\caption{Brain Tumor}
\label{fig:task01_dims}
\end{subfigure}
\hfill
\begin{subfigure}[b]{0.49\linewidth}
\centering
\includegraphics[width=\linewidth]{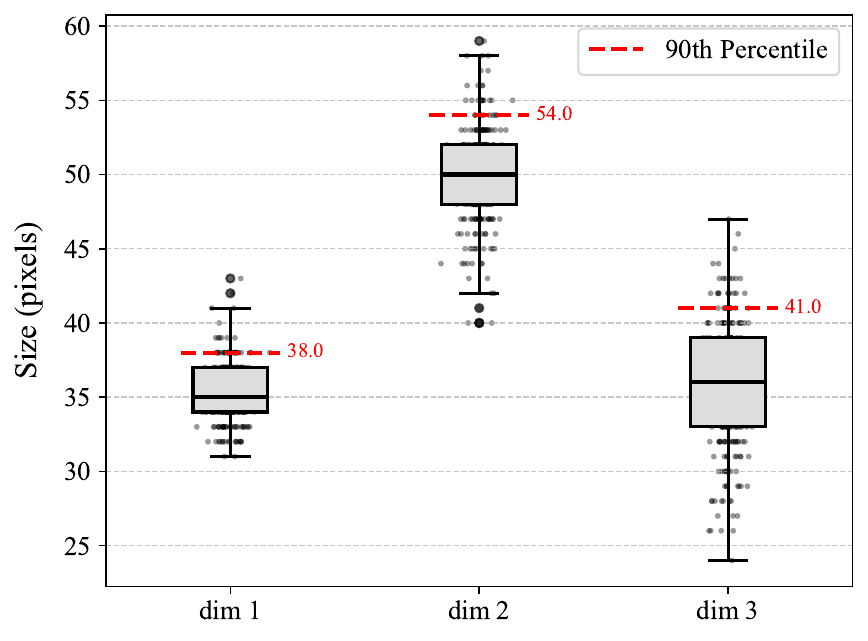}
\caption{Hippocampus}
\label{fig:task04_dims}
\end{subfigure}
\caption{Distribution of spatial dimensions across the two datasets. Dashed lines indicate the pre-rounded 90\textsuperscript{th} percentile values used to define the standardized input sizes for training.}
\label{fig:all_task_dims}
\end{figure}

This task involves segmenting three tumor subregions: enhancing tumor (ET), peritumoral edema (PE), and necrotic core (NC), from multi-modal MRI scans. Each volume includes four sequences: FLAIR, T1, T1Gd, and T2. The dataset comprises 484 labeled 4D volumes, which we split into 80\% for training and validation (387 volumes) and 20\% for testing (97 volumes). Following the same input standardization protocol, we computed the 90\textsuperscript{th} percentile of each spatial axis (see Fig. \ref{fig:all_task_dims}) over the training set and rounded up to the nearest power of two, yielding a fixed input size of 256$\times$256$\times$128 with four channels. All images were then padded or cropped accordingly.

\paragraph{\textbf{Hippocampus.}}

This task focuses on segmenting the left (L) and right (R) hippocampus from the background using T1-weighted MRI scans. The dataset includes 260 labeled 3D volumes. We randomly split the data into 80\% for training and validation (208 volumes) and 20\% for testing (52 volumes). To standardize the input size for semantic segmentation, we computed the 90\textsuperscript{th} percentile of each spatial dimension (see Fig. \ref{fig:all_task_dims}) across the training set and rounded up to the nearest power of two, resulting in a final input size of 32$\times$64$\times$32. All images were padded or cropped to match this size. Table \ref{tab:datasets} summarizes the key statistics of the brain imaging datasets used in our experiments.
 
\begin{table}[!t]
\centering
\caption{Summary statistics of the brain imaging datasets used in this study.}
\vspace{0.1cm}
\begin{tabular}{l@{\hspace{8pt}}c@{\hspace{8pt}}c@{\hspace{8pt}}c@{\hspace{8pt}}c@{\hspace{8pt}}r}
\toprule
\textbf{Task} & \textbf{Modalities} & \textbf{\#Classes} & \textbf{\#Volumes} & \textbf{Input size} \\
\midrule
\textbf{Hippocampus} & T1 & 3 & 260 & 32$\times$64$\times$32 \\
\textbf{Brain Tumor} & FLAIR, T1, T1Gd, T2 & 4 & 484 & 256$\times$256$\times$128 \\ 
\bottomrule
\end{tabular}
\label{tab:datasets}
\end{table}

\section{Detailed Results}

\begin{table}[!t]
\centering
\footnotesize
\caption{Mean Dice Similarity Coefficient (DSC) per class across all scales and model families for Hippocampus (HP) and Brain Tumor (BT) test datasets.}
\vspace{0.1cm}
\label{tab:mean-dsc-all-cls}
\resizebox{\textwidth}{!}{
\begin{tabular}{lccccc}
\toprule
\textbf{Method (Scale)} & \textbf{HP--L} & \textbf{HP--R} & \textbf{BT--PE} & \textbf{BT--NC} & \textbf{BT--ET} \\
\midrule
UNet (1) & \underline{0.886}$\pm$0.031 & \underline{0.862}$\pm$0.050 & \underline{0.788}$\pm$0.121 & 0.581$\pm$0.236 & \underline{0.728}$\pm$0.256 \\
UNet+Aug (1) & 0.879$\pm$0.033 & \underline{0.862}$\pm$0.038 & 0.773$\pm$0.126 & 0.566$\pm$0.222 & 0.725$\pm$0.248 \\
nnUNet (1) & \textbf{0.889}$\pm$0.033 & \textbf{0.871}$\pm$0.034 & \textbf{0.794}$\pm$0.112 & \textbf{0.591}$\pm$0.253 & \textbf{0.752}$\pm$0.243 \\
RARE-UNet (1) & 0.878$\pm$0.032 & 0.860$\pm$0.037 & 0.778$\pm$0.128 & \underline{0.586}$\pm$0.230 & 0.717$\pm$0.263 \\
\midrule
UNet--Pad (1/2) & 0.013$\pm$0.036 & 0.056$\pm$0.058 & 0.332$\pm$0.226 & 0.253$\pm$0.246 & 0.041$\pm$0.050 \\
UNet--Up (1/2) & 0.864$\pm$0.046 & 0.831$\pm$0.079 & \underline{0.771}$\pm$0.124 & \underline{0.547}$\pm$0.243 & \underline{0.714}$\pm$0.249 \\
UNet+Aug--Pad (1/2) & 0.258$\pm$0.050 & 0.168$\pm$0.040 & 0.071$\pm$0.054 & 0.058$\pm$0.107 & 0.089$\pm$0.125 \\
UNet+Aug--Up (1/2) & \underline{0.866}$\pm$0.040 & \underline{0.854}$\pm$0.045 & 0.764$\pm$0.127 & 0.536$\pm$0.236 & \textbf{0.720}$\pm$0.235 \\
nnUNet--Pad (1/2) & 0.835$\pm$0.038 & 0.804$\pm$0.063 & 0.765$\pm$0.120 & 0.537$\pm$0.256 & 0.706$\pm$0.248 \\
nnUNet--Up (1/2) & 0.831$\pm$0.066 & 0.808$\pm$0.075 & 0.750$\pm$0.127 & 0.518$\pm$0.255 & 0.711$\pm$0.250 \\
RARE-UNet (1/2) & \textbf{0.868}$\pm$0.041 & \textbf{0.856}$\pm$0.042 & \textbf{0.788}$\pm$0.120 & \textbf{0.587}$\pm$0.231 & 0.706$\pm$0.266 \\
\midrule
UNet--Pad (1/4) & 0.000$\pm$0.000 & 0.000$\pm$0.000 & 0.276$\pm$0.194 & 0.220$\pm$0.239 & 0.032$\pm$0.032 \\
UNet--Up (1/4) & 0.788$\pm$0.071 & 0.765$\pm$0.118 & 0.732$\pm$0.127 & 0.466$\pm$0.244 & 0.663$\pm$0.242 \\
UNet+Aug--Pad (1/4) & 0.104$\pm$0.044 & 0.028$\pm$0.023 & 0.006$\pm$0.011 & 0.040$\pm$0.082 & 0.052$\pm$0.070 \\
UNet+Aug--Up (1/4) & \underline{0.831}$\pm$0.087 & \underline{0.813}$\pm$0.073 & \underline{0.736}$\pm$0.128 & 0.464$\pm$0.236 & \textbf{0.680}$\pm$0.219 \\
nnUNet--Pad (1/4) & 0.552$\pm$0.111 & 0.505$\pm$0.113 & 0.682$\pm$0.133 & \underline{0.468}$\pm$0.234 & 0.598$\pm$0.230 \\
nnUNet--Up (1/4) & 0.586$\pm$0.135 & 0.609$\pm$0.121 & 0.698$\pm$0.137 & 0.444$\pm$0.244 & 0.644$\pm$0.253 \\
RARE-UNet (1/4) & \textbf{0.846}$\pm$0.077 & \textbf{0.824}$\pm$0.064 & \textbf{0.760}$\pm$0.120 & \textbf{0.524}$\pm$0.247 & \underline{0.671}$\pm$0.248 \\
\midrule
UNet--Pad (1/8) & 0.000$\pm$0.000 & 0.000$\pm$0.000 & 0.098$\pm$0.119 & 0.076$\pm$0.138 & 0.024$\pm$0.026 \\
UNet--Up (1/8) & 0.536$\pm$0.233 & 0.132$\pm$0.198 & 0.619$\pm$0.146 & 0.367$\pm$0.264 & 0.494$\pm$0.286 \\
UNet+Aug--Pad (1/8) & 0.000$\pm$0.000 & 0.000$\pm$0.000 & 0.002$\pm$0.012 & 0.035$\pm$0.174 & 0.100$\pm$0.231 \\
UNet+Aug--Up (1/8) & \textbf{0.793}$\pm$0.153 & \underline{0.785}$\pm$0.193 & \underline{0.637}$\pm$0.139 & \underline{0.376}$\pm$0.258 & \underline{0.513}$\pm$0.267 \\
nnUNet--Pad (1/8) & 0.054$\pm$0.120 & 0.005$\pm$0.039 & 0.491$\pm$0.169 & 0.330$\pm$0.211 & 0.377$\pm$0.237 \\
nnUNet--Up (1/8) & 0.087$\pm$0.135 & 0.023$\pm$0.081 & 0.553$\pm$0.185 & 0.321$\pm$0.233 & 0.482$\pm$0.288 \\
RARE-UNet (1/8) & \underline{0.779}$\pm$0.195 & \textbf{0.791}$\pm$0.177 & \textbf{0.668}$\pm$0.146 & \textbf{0.463}$\pm$0.263 & \textbf{0.570}$\pm$0.283 \\
\midrule
UNet--Pad (Overall) & 0.225$\pm$0.382 & 0.229$\pm$0.366 & 0.374$\pm$0.254 & 0.282$\pm$0.185 & 0.206$\pm$0.302 \\
UNet--Up (Overall) & 0.768$\pm$0.139 & 0.648$\pm$0.300 & 0.727$\pm$0.066 & \underline{0.490}$\pm$0.083 & 0.650$\pm$0.093 \\
UNet+Aug--Pad (Overall) & 0.310$\pm$0.341 & 0.265$\pm$0.351 & 0.213$\pm$0.324 & 0.175$\pm$0.226 & 0.241$\pm$0.280 \\
UNet+Aug--Up (Overall) & \underline{0.842}$\pm$0.033 & \underline{0.828}$\pm$0.031 & \underline{0.728}$\pm$0.054 & 0.486$\pm$0.073 & \underline{0.659}$\pm$0.086 \\
nnUNet--Pad (Overall) & 0.582$\pm$0.331 & 0.546$\pm$0.341 & 0.683$\pm$0.119 & 0.482$\pm$0.098 & 0.608$\pm$0.145 \\
nnUNet--Up (Overall) & 0.598$\pm$0.316 & 0.578$\pm$0.334 & 0.699$\pm$0.091 & 0.469$\pm$0.100 & 0.647$\pm$0.103 \\
RARE-UNet (Overall) & \textbf{0.843}$\pm$0.039 & \textbf{0.833}$\pm$0.028 & \textbf{0.748}$\pm$0.047 & \textbf{0.540}$\pm$0.051 & \textbf{0.666}$\pm$0.058 \\
\bottomrule
\end{tabular}
}
\end{table}

Table \ref{tab:mean-dsc-all-cls} presents the mean DSCs for individual classes at different resolutions in the hippocampus dataset (Class 1 and Class 2, representing the left and right hippocampus regions) and the brain tumor dataset: Class 1 (Edema), Class 2 (Non-enhancing tumor), and Class 3 (Enhancing tumor). The results suggest that the left hippocampus is slightly easier to segment, but overall, the DSC scores for the two classes are closely aligned, indicating balanced model performance across hippocampus substructures. Moreover, Class 2 in brain tumors proves to be the most challenging to segment.

\end{document}